\documentstyle[11pt,epsfig,aaspp4]{article}

\begin{document}

\title{GRB 030226 in a Density-Jump Medium}
\author{Z. G. Dai and X. F. Wu}
\affil{Department of Astronomy, Nanjing University, Nanjing 210093, China\\
daizigao@public1.ptt.js.cn, xfwu@nju.edu.cn}

\begin{abstract}
We present an explanation for the unusual temporal feature of the
GRB 030226 afterglow. The R-band afterglow of this burst faded as
$\sim t^{-1.2}$ in $\sim 0.2$ days after the burst, rebrightened
during the period of $\sim 0.2-0.5$ days, and then declined with
$\sim t^{-2.0}$. To fit such a light curve, we consider an
ultrarelativistic jetted blast wave expanding in a density-jump
medium. The interaction of the blast wave with a large density
jump produces relativistic reverse and forward shocks. In this
model, the observed rebrightening is due to emissions from these
newly forming shocks, and the late-time afterglow is caused by
sideways expansion of the jet. Our fitting implies that the
progenitor star of GRB 030226 could have produced a stellar wind
with a large density jump prior to the GRB onset.

\end{abstract}


\section {Introduction}

The gamma-ray burst (GRB) 030226 was detected by the {\em High
Energy Transient Explorer 2} Satellite ({\em HETE-2}) at
03:46:31.99 UT on 26 February 2003 (Suzuki et al. 2003). The burst
duration in the 30-400 keV band was beyond 100 seconds, and the
fluence in this energy band was $\sim 5.7\times 10^{-6}\,{\rm
ergs}\,{\rm cm}^{-2}$. Owing to the rapid localization of {\em HETE-2},
an early afterglow of this burst was observed. Fox, Chen \& Price (2003)
detected an optical counterpart $\sim 0.11$ days after the burst, and
Price, Fox \& Chen (2003) found about 9 minutes later that it had faded.
The subsequent spectroscopic observations revealed some absorption lines,
indicating that this burst was at a redshift of $z=1.986 \pm 0.001$
(Greiner et al. 2003a). Thus, the isotropic $\gamma$-ray energy
release is estimated to be $\sim 6\times 10^{52}$ ergs, assuming a
cosmological model with $\Omega_m=0.3$, $\Omega_\Lambda=0.7$
and $H_0=65\,{\rm km}\,{\rm s}^{-1}\,{\rm Mpc}^{-1}$.

The afterglow light curve of GRB 030226 has one unusual feature.
The R-band afterglow declined as $\sim t^{-1.2}$ shown by the four
observations during the period of $\sim 0.13-0.20$ days after the
burst, and about seven hours later, it began to decay with another
simple power-law of $t^{-2.0}$ as late-time afterglows from jets
usually do. However, contrary to jetted afterglows observed
previously, the extrapolation of the early afterglow light curve
of GRB 030226 crosses with the late-time light curve at $\sim 2$
days rather than $\sim 0.5$ days, which implies an obvious
rebrightening. Such a feature could result from post-burst energy 
injection. A few variants of this model have been studied (e.g., Dai 
\& Lu 1998a; Panaitescu, M\'esz\'aros \& Rees 1998; Zhang \& 
M\'esz\'aros 2002; Nakar, Piran \& Granot 2003). In this Letter, 
we explain the GRB 030226 afterglow light curve by
assuming this burst to be in a density-jump medium (see \S 2). In \S 3 
we discuss scenarios for density-jump formation and give
order-of-magnitude estimates of the density-jump factor,  which are
consistent with required by our fitting. Our results are summarized in \S 4.

\section{The Model}

Before proposing our model, we first discuss implications of the
observations on the GRB 030226 afterglow: The observed light curve
index ($\alpha_2$) after $\sim 0.5$ days since the {\em HETE-2}
trigger suggests that this afterglow might have originated from a
highly collimated relativistic shock, because the transition of
this shock from the spherical to spreading phase could lead to
steepening of the light curve to the flux proportional to $t^{-p}$
for $p>2$, where $p$ is the spectral index of the
shock-accelerated electrons (Rhoads 1999; Sari, Piran \& Halpern
1999; for the case of $1<p<2$ see Dai \& Cheng 2001). Thus,
$p=\alpha_2\simeq 2$ for GRB 030226. If the early afterglow is
assumed to result from the shock expanding in a medium with
density of $n_w\propto R^{-k}$, then the light curve index in the
slow-cooling regime becomes $\alpha_1=3(p-1)/4+k/(8-2k)$ (Dai \&
Lu 1998b; M\'esz\'aros, Rees \& Wijers 1998). The observed value
of $\alpha_1\simeq 1.2$, combined with the inferred value of $p$,
implies $k\simeq 2$, revealing a wind surrounding the burst. This
environment as an expected signature of massive progenitor stars
of GRBs was discussed theoretically in detail (Dai \& Lu 1998b;
M\'esz\'aros et al. 1998; Chevalier \& Li 1999, 2000) and
confirmed observationally for some bursts (see, e.g., Price et al.
2002 for GRB 011121). Recently, the wind interaction model was
also shown to explain well the early afterglow of GRB 021004 (Li
\& Chevalier 2003).

Therefore, we consider an ultrarelativistic jet with an isotropic
energy of $E$, which first expands in a circumburst wind of the
density profile of $n_w(R)=AR^{-2}$ with $A=3\times
10^{35}A_*\,{\rm cm}^{-1}$ for $R<R_0$. Here $A_*=({\dot
M}/10^{-5}M_\odot\, {\rm yr}^{-1})/(V_w/10^3{\rm km}\,{\rm
s}^{-1})$, where ${\dot M}$ and $V_w$ are the mass loss rate and
velocity of the wind respectively. After the internal shock
emission, the jet will start to sweep up its ambient wind, leading
to an ultrarelativistic blast wave. The initial hydrodynamics and
emission were recently discussed by Wu et al. (2003) and Kobayashi
\& Zhang (2003). We further assume this blast wave which will hit
an outer high-density homogeneous region for $R\ge R_0$. Dai \& Lu
(2002, hereafter DL02) have analyzed the hydrodynamics and
afterglow emission from a relativistic blast wave in detail when
it expands in a density-jump medium, which might have been
produced around $\sim 10^{18}$ cm due to the interaction of a
stellar wind with the ambient matter (see \S 3).

The interaction of a relativistic blast wave with a large density
jump can be described through two shocks: a reverse shock that
propagates into the hot shell (viz., the $R\le R_0$ wind matter
swept up by the blast wave), and a forward shock that propagates
into the outer high-density medium. Thus, there are four regions
separated in the system of interest by the two shocks: (1) the
unshocked high-density medium, (2) the forward-shocked
high-density medium, (3) the reverse-shocked hot shell, and (4)
the unshocked hot shell. We denote $\gamma_i$ as the Lorentz
factor of region ``$i$" measured in the local medium's rest frame,
and $n_1$ is the baryon number density of the medium at $R\ge R_0$
(region 1). Defining the ratio ($f$) of the energy densities of
regions 4 and 1, DL02 have got
\begin{equation}
\gamma_2=\gamma_3=\frac{\gamma_4^{1/2}f^{1/4}}{3^{1/4}},
\end{equation}
and found that a density jump with a factor of much more than $21$
can lead to a relativistic reverse shock. In this case, the Lorentz factor 
of regions 2 and 3 can further be scaled as
$\gamma_2=\gamma_3=1.4(1+z)^{1/2}A_*^{1/4}n_{1,3}^{-1/4}
t_0^{-1/2}(R/R_0)^{-1}$,
where $n_{1,3}=n_1/10^3 \,{\rm cm}^{-3}$, and $t_0$ is the
observed time (in 1 day) at $R=R_0=8.0\times
10^{17}(1+z)^{-1/2}E_{53}^{1/2}A_*^{-1/2}t_0^{1/2}\,{\rm cm}$
(where $E_{53}$ is the isotropic-equivalent energy of the jet in
units of $10^{53}$ ergs). Comparing $\gamma_2$ (or $\gamma_3$) 
with the Lorentz factor of region 4 at the crossing time,
$\gamma_4=8.8(1+z)^{1/4}E_{53}^{1/4} A_*^{-1/4}t_0^{-1/4}$, it is
easy to see that the Lorentz factor of region 3 is much less than
that of region 4 for typical parameters, showing that most of the
initial kinetic energy is converted into thermal energy by the
shocks. The radius at which the reverse shock has just crossed
region 4 is derived as $R_\Delta=R_0[1+1.25\times 10^{-2}
(1+z)^{1/2}E_{53}^{-1/2}A_*n_{1,3}^{-1/2}t_0^{-1/2}]$ (DL02).
Thus, $R_\Delta/R_0-1\ll 1$ for typical parameters, so that the
Lorentz factor of region 2 at $R=R_\Delta$ is well approximated by
its Lorentz factor at $R=R_0$. If the initial half opening angle
of the jet is $\theta_0\sim 0.1$ as in most of the afterglows
analyzed by Frail et al. (2001), then it is possible that after
the crossing time the Lorentz factor of the jet is far below
$\theta_0^{-1}$ and therefore the flux decays as $t^{-p}$ due to
sideways expansion as long as the jet is still relativistic
(Rhoads 1999; Sari et al. 1999).

In the standard afterglow model (Sari, Piran \& Narayan 1998), the
synchrotron flux $F_{\nu,i}$ of region ``$i$" in optical to X-ray
bands is determined by three quantities: the typical synchrotron
frequency $\nu_{m,i}$, the cooling frequency $\nu_{c,i}$, and the
peak flux $F_{\nu, max,i}$ . We divide the whole evolution into
three stages: (I) before, (II) during, and (III) after  the
crossing time. Using Chevalier \& Li's (2000) results, we get
these three quantities at stage I: $\nu_m^{\rm I}(t)=4.8\times
10^{13}(1+z)^{1/2}
\epsilon_{e,-1}^2\epsilon_{B,-1}^{1/2}\xi^2E_{53}^{1/2}t^{-3/2}$\,Hz,
$\nu_c^{\rm I}(t)=6.6\times
10^{13}(1+z)^{-3/2}\epsilon_{B,-1}^{-3/2}
E_{53}^{1/2}A_*^{-2}t^{1/2}$\,Hz, and $F_{\nu, max}^{\rm
I}(t)=0.59(1+z)^{3/2}
\epsilon_{B,-1}^{1/2}E_{53}^{1/2}A_*t^{-1/2}D_{L,28}^{-2}$\,Jy,
where $\xi=(p-2)/(p-1)$, $\epsilon_e=0.1\epsilon_{e,-1}$ and
$\epsilon_B =0.1\epsilon_{B,-1}$ are constant fractions of the
internal energy density going into the electrons and the magnetic
fields respectively, $t$ is the observer's time in units of 1 day,
and $D_{L,28}$ is the luminosity distance to the source in units
of $10^{28}$ cm.

At stage II,  the spectral break frequencies and the peak flux
evolve as
\begin{equation}
\nu_{m,i}^{\rm II} = \left \{
   \begin{array}{lll}
   1.3\times 10^{12}(1+z)\epsilon_{e,-1}^2\epsilon_{B,-1}^{1/2}\xi^2A_*
       n_{1,3}^{-1/2} t_0^{-2}(R/R_0)^{-4}{\rm Hz},   & {\rm if}\,\,i=2,\\
   1.0\times 10^{15}\epsilon_{e,-1}^2\epsilon_{B,-1}^{1/2}\xi^2E_{53}
       A_*^{-1}n_{1,3}^{1/2}t_0^{-1}(R/R_0)^{-2}{\rm Hz},   & {\rm if}\,\,i=3,\\
   \nu_m^{\rm I}(t_0)(R/R_0)^{-4}, & {\rm if}\,\,i=4.
        \end{array}
       \right.
\end{equation}
\begin{equation}
\nu_{c,i}^{\rm II} = \left \{
   \begin{array}{ll}
   1.0\times 10^{12}(1+z)^{-1}\epsilon_{B,-1}^{-3/2}A_*^{-1}n_{1,3}^{-1/2}
       (t/t_0)^{-2}(R/R_0)^4{\rm Hz}, & {\rm if}\,\,i=2,3,\\
   \nu_c^{\rm I}(t_0)(R/R_0)^{-4}, & {\rm if}\,\,i=4.
       \end{array}
       \right.
\end{equation}
\begin{equation}
F_{\nu,max,i}^{\rm II} = \left \{
   \begin{array}{lll}
   470(1+z)^{1/2}\epsilon_{B,-1}^{1/2}E_{53}^{1/2}A_*^{-1}n_{1,3}
       t_0^{1/2}D_{L,28}^{-2}[(R/R_0)^3-1](R/R_0)^{-2}\,{\rm Jy}, & {\rm if}\,\,i=2,\\
   28(1+z)\epsilon_{B,-1}^{1/2}E_{53}n_{1,3}^{1/2}D_{L,28}^{-2}
       [(R/R_0)^2-1](R/R_0)^{-2}\,{\rm Jy},   & {\rm if}\,\,i=3,\\
   F_{\nu,max}^{\rm I}(t_0)\zeta (R/R_0)^{-2}, & {\rm if}\,\,i=4,
        \end{array}
       \right.
\end{equation}
where $\zeta$ is the ratio of the total electron numbers of region
4 at radius $R\in (R_0,R_\Delta)$ and at radius $R_0$. It should
be emphasized that $\nu_{c,4}^{\rm II}$ in equation (3) is in fact
a cutoff frequency because of adiabatic expansion of region 4,
implying that there is no emission from this region at frequencies
above $\nu_{c,4}^{\rm II}$

From DL02, we obtain the time at which $R=R_\Delta$ as follows:
$t_{\Delta,2} =t_{\Delta,3}=2.0t_0$ for regions 2 and 3. If the
afterglow originates from a jet, one expects that at  stage III
the break frequencies and the peak flux of region 2 are given by
$\nu_{m,2}^{\rm III}=\nu_{m,2}^{\rm II}
(t_{\Delta,2})(t/t_{\Delta,2})^{-2}$, $\nu_{c,2}^{\rm
III}=\nu_{c,2}^{\rm II} (t_{\Delta,2})$, and $F_{\nu,max,2}^{\rm
III}=F_{\nu,max,2}^{\rm II} (t_{\Delta,2})(t/t_{\Delta,2})^{-1}$
(Sari et al. 1999). For the parameters adopted below, the cutoff
frequency of region 3 is less than the optical frequency so that
there is no emission from this region at frequencies above the
optical band.

Figure 1 presents an example case of our fitting to the
multiwavelength light curves of the GRB 030226 afterglow. Please
note that an error of 0.05 mag for the initial four afterglow data
points in the figure has been taken as the error in the observed
brightness of two calibration stars by Garnavich, von Braun \&
Stanek (2003). This error is smaller than a conservative estimate
of 0.3 mag by Zeh et al. (2003). The model parameters are taken:
$\epsilon_{e,-1}=0.9$, $\epsilon_{B,-1}=4$, $E_{53}=1.0$,
$A_*=0.01$, $p=2.01$, $t_0=0.22$ days, and $n_{1,3}=0.1$. If the
emission off the line of sight (LOS) were neglected, one would see
an initial drop at $t\simeq t_0$, a rapid brightening at
$t_0\lesssim t\lesssim t_{\Delta,2}$ and an abrupt decay at
$t\simeq t_{\Delta,2}$ for all the light curves (DL02). However,
such features are expected to be smoothed by the off-LOS emission.
In Figure 1, we have considered the off-LOS emission effect for
region 4 due to its ultrarelativistic expansion (e.g.,
$\gamma_4\sim 53 \gg 1$ for the taken model parameters) at
$t\simeq t_0$, following Kumar \& Panaitescu (2000). However, this
effect is insignificant for region 3 because $\gamma_3\sim 2.9$ is
not far greater than unity at $t\simeq t_{\Delta,2}$. 

We next discuss implications of the fitting. First, because the
late-time afterglow at high frequencies in our model results from
fast-cooling electrons, the spectral index should be
$\beta=p/2\simeq 1.0$, which is consistent with the observed
values, $\beta_{\rm opt,\, ob}=0.99\pm 0.42$ at the optical band
(Nysewander et al. 2003) and $\beta_{\rm X,\, ob}=1.0\pm 0.1$ at
the X-ray band (Pedersen et al. 2003; Sako \& Fox 2003),
respectively. Second, the jet's isotropic-equivalent energy ($\sim
10^{53}$ ergs) from our fitting is close to  the isotropic
$\gamma$-ray energy release ($\sim 6\times 10^{52}$ ergs). This
shows that the efficiency of $\gamma$-ray emission from internal
shocks is as high as $\sim 40\%$. Third, our fitting gives a weak
wind $A_*\sim 0.01$, which implies non-detection of an optical
flash from GRB 030226, as in most of the GRBs observed previously
(Wu et al. 2003). Finally, the required value of $n_1$ indicates
that the progenitor star of GRB 030226 could have been in a dense
medium such as a giant molecular cloud (GMC).

\section{Density-Jump Formation}

In this section we show that a large density jump seems to form in
the vicinity of a massive progenitor star. A natural scenario for
density-jump formation is the interaction of a wind from a massive
star with its outer environment. This interaction produces a
stellar wind bubble. Numerical simulations of the bubble dynamics
in a homogeneous interstellar medium were performed by
Ramirez-Ruiz et al. (2001), and here we give analytical studies to
obtain the order-of-magnitude factor of a density jump by
considering two kinds of environment. First, we discuss a simple
case in which the outer environment of a massive star is a
homogeneous GMC with density $n_1$. Owing to the presence of two
shocks (viz., a reverse shock that propagates into the wind gas
and a forward shock that propagates into the GMC gas), the system
has a four-zone structure consisting of (1) the unshocked GMC gas,
(2) the shocked GMC shell, (3) the shocked wind gas, and (4) the
unshocked wind gas moving with a velocity $V_w$ and with a mass
loss rate $\dot{M}$, where zones 2 and 3 are separated by a
contact discontinuity. Assuming an adiabatic expansion and two
strong shocks (Castor, McCray \& Weaver 1975; Wijers 2001; Mirabal
et al. 2003), we get the forward shock radius
\begin{equation}
R_{\rm sh}^{\rm f}\simeq 4.8\times 10^{18}{\dot
M}_{-7}^{1/5}V_{w,3}^{2/5}n_{1,2}^{-1/5}t_{w,5}^{3/5}\,{\rm cm},
\end{equation}
and the reverse shock radius
\begin{equation}
R_{\rm sh}^{\rm r}\simeq 7.0\times 10^{17}{\dot
M}_{-7}^{3/10}V_{w,3}^{1/10}n_{1,2}^{-3/10}t_{w,5}^{2/5}\,{\rm
cm},
\end{equation}
where ${\dot M}_{-7}={\dot M}/10^{-7}M_\odot\,{\rm yr}^{-1}$,
$V_{w,3}=V_w/10^3{\rm km}\,{\rm s}^{-1}$, $n_{1,2}=n_1/10^2\,{\rm
cm}^{-3}$, and $t_{w,5}$ is the wind lifetime in units of $10^5$
years. Here the reason for scaling the stellar mass loss rate with
$10^{-7}M_\odot\,{\rm yr}^{-1}$ is that $A_*\sim 0.01$ in our
fitting implies ${\dot M}\sim 10^{-7}V_{w,3}M_\odot\,{\rm
yr}^{-1}$. This scaling is typical for main-sequence stellar
winds, and further not far below typical mass loss rates of
Wolf-Rayet stars (Willis 1991). One large density jump appears at
the contact discontinuity, whose factor is approximated by
\begin{equation}
\chi=\frac{n_c}{n_w(R_{\rm sh}^{\rm r})}\sim 1.7\times 10^4{\dot
M}_{-7}^{-2/5}V_{w,3}^{6/5}n_{1,2}^{2/5}t_{w,5}^{4/5}.
\end{equation}
The radius at which the density jump is required to appear in our
fitting is estimated as $R_0\sim 2.2\times 10^{18}$ cm, which is
larger than $R_{\rm sh}^{\rm r}$ and closer to $R_{\rm sh}^{\rm
f}$ for reasonable wind parameters, so the afterglow shock wave
would be able to reach the density jump at the early time when the
rebrightening was observed in the GRB 030226 afterglow light
curve.

Second, the surrounding medium of a fast wind from a massive star
(e.g., a Wolf-Rayet star) may be a dense slow wind, because this
star could have passed through a red supergiant (RSG) phase.
During such a phase, the star has produced a wind with a larger
mass loss rate ${\dot M}_{\rm RSG}$ and a slower velocity $V_{\rm
RSG}$. This wind has formed a medium with density $n_{\rm
RSG}(R)=3\times 10^{37}({\dot M}_{{\rm RSG},-5}/V_{{\rm
RSG},1})R^{-2}\,{\rm cm}^{-1}$, where ${\dot M}_{{\rm
RSG},-5}={\dot M}_{\rm RSG}/10^{-5}M_\odot\,{\rm yr}^{-1}$ and
$V_{{\rm RSG},1}=V_{\rm RSG}/10\,{\rm km}\,{\rm s}^{-1}$. The
interaction of a RSG wind with its homogeneous environment (e.g.,
a GMC) could have given rise to a RSG wind bubble. The dynamics of
this bubble is similar to the case discussed above, provided that
they are both adiabatic. From equation (5), we can write directly
the radius of a thin shell swept up by the RSG wind as
$R_{\rm sh, RSG}\simeq 1.9\times 10^{18}{\dot M}_{{\rm RSG},
-5}^{1/5}V_{{\rm RSG},1}^{2/5}n_{1,2}^{-1/5}t_{{\rm
RSG},5}^{3/5}\,{\rm cm}$,
where $t_{{\rm RSG},5}$ is the RSG wind lifetime in units of
$10^5$ years. Also, we obtain the factor of a density jump
appearing at the contact discontinuity in the RSG wind bubble,
$\chi_{\rm RSG} \sim 11{\dot M}_{{\rm RSG}, -5}^{-2/5}V_{{\rm
RSG},1}^{6/5}n_{1,2}^{2/5}t_{{\rm RSG},5}^{4/5}$.

Furthermore, a fast wind from the Wolf-Rayet phase sweeps up the
pre-existing slow wind, leading possibly to another wind bubble.
Chevalier \& Imamura (1983) gave self-similar solutions of this
bubble. Assuming an adiabatic evolution and two strong shocks 
(Garcia-Segura \& Mac Low 1995), we have the forward shock radius
\begin{equation}
R_{\rm sh}^{\rm f}\simeq 8.6\times 10^{18}{\dot
M}_{-7}^{1/3}V_{w,3}^{2/3}{\dot M}_{{\rm RSG}, -5}^{-1/3}V_{{\rm
RSG},1}^{1/3}t_{w,5}\,{\rm cm},
\end{equation}
and the reverse shock radius
\begin{equation}
R_{\rm sh}^{\rm r}\simeq 1.6\times 10^{18}{\dot
M}_{-7}^{1/2}V_{w,3}^{1/2}{\dot M}_{{\rm RSG}, -5}^{-1/2}V_{{\rm
RSG},1}^{1/2}t_{w,5}\,{\rm cm}.
\end{equation}
Letting $R_{\rm sh}^{\rm f}$ in equation (8) be equal to $R_{\rm sh, RSG}$, 
we define a critical lifetime of the Wolf-Rayet wind, 
\begin{equation}
t_w^{\rm cr}\simeq 2.2\times 10^4{\dot
M}_{-7}^{-1/3}V_{w,3}^{-2/3}{\dot M}_{{\rm RSG}, -5}^{8/15}V_{{\rm
RSG},1}^{1/15}n_{1,2}^{-1/5}t_{{\rm RSG},5}^{3/5}\,{\rm yrs}.
\end{equation}
If $t_w\lesssim t_w^{\rm cr}$, the Wolf-Rayet wind only sweeps up
the RSG wind. In this subcase, a large density jump occurs at the
contact discontinuity and its factor is $\chi_w=n_{\rm RSG}(R_{\rm
sh}^{\rm f})/n_w(R_{\rm sh}^{\rm r})\sim 360{\dot
M}_{-7}^{-2/3}V_{w,3}^{2/3}{\dot M}_{{\rm RSG}, -5}^{2/3}V_{{\rm
RSG},1}^{-2/3}$, where $R_{\rm sh}^{\rm f}$ and $R_{\rm sh}^{\rm
r}$ are given by equations (8) and (9) respectively. When
$R_{\rm sh}^{\rm f}\sim R_{\rm sh, RSG}$, the above two density
jumps merge to one larger density jump with a factor of   
\begin{equation}
\chi\sim \chi_w\chi_{\rm RSG} \sim  4\times 10^3{\dot
M}_{-7}^{-2/3}V_{w,3}^{2/3}{\dot M}_{{\rm RSG}, -5}^{4/15}V_{{\rm
RSG},1}^{8/15}n_{1,2}^{2/5}t_{{\rm RSG},5}^{4/5}.
\end{equation}
On the other hand, if $t_w> t_w^{\rm cr}$, the Wolf-Rayet wind not
only sweeps up the RSG wind but also the outer GMC gas, in which
subcase the dynamics can be approximated by equations (5) and (6),
and thus the factor of a large density jump that occurs at the
contact discontinuity is estimated by equation (7). Because $R_0$
is larger than $R_{\rm sh}^{\rm r}$ and closer to $R_{\rm sh}^{\rm
f}$ in both subcases, the afterglow shock wave would be able to
arrive at the large density jump at an expected early time of GRB
030226.

\section{Conclusions}

In this paper we have shown that density jumps (or bumps) are
likely to occur in the vicinity of a GRB, especially if a stellar
wind has interacted with the ambient matter or if winds ejected at
different velocities have collided each other, prior to the GRB
onset. We have considered an ultrarelativistic {\em jetted} blast
wave in a density-jump medium to explain the unusual temporal
feature of the GRB 030226 afterglow. A large density jump can
produce a relativistic reverse shock, which leads naturally to (1)
effective conversion of the initial kinetic energy to thermal
energy and (2) rapid deceleration of the jet. Therefore, the
assumption of one large density jump has two advantages. Based on
them, the observed rebrightening is understood as due to emissions
from reverse and forward shocks forming in the interaction of the
blast wave with the density jump, and the late-time afterglow is
caused by sideways expansion of the jet.

 \acknowledgments
We thank the referee and X. Y. Wang for valuable comments. This
work was supported by the National Natural Science Foundation of
China (grant 10233010) and the National 973 Project (NKBRSF
G19990754).

\clearpage
\begin{figure}
\begin{picture}(100,250)
\put(0,0){\includegraphics{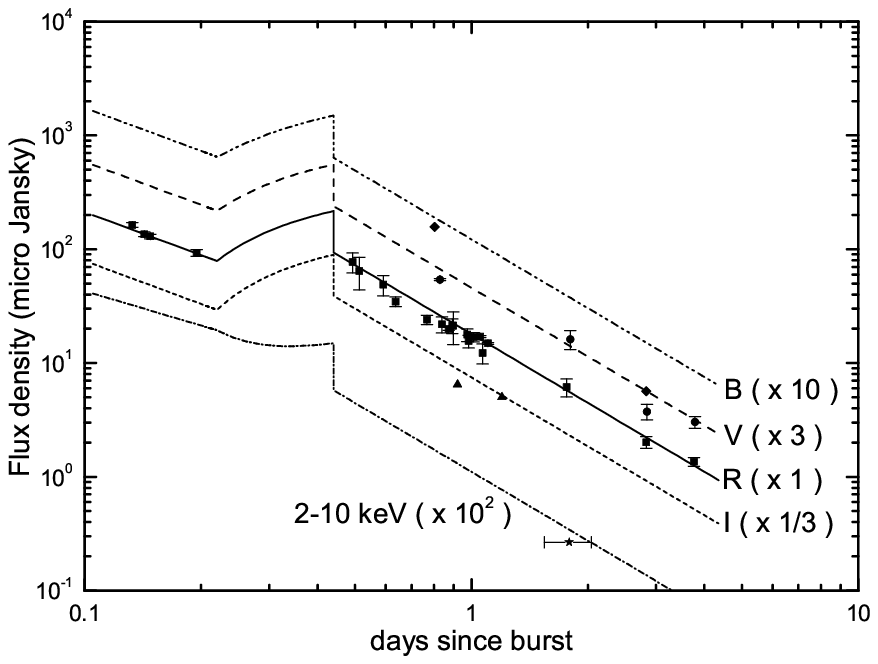}}
\end{picture}
\caption {Fitting of the multiwavelength afterglow of GRB 030226
by assuming an ultrarelativistic blast wave which expands in a
density-jump medium. The model parameters are seen in the text.
Data are from Ando et al. 2003a, 2003b; Covino et al. 2003;
Fatkhullin et al. 2003; Garnavich, von Braun \& Stanek 2003;
Greiner et al. 2003b;Guarnieri et al. 2003; Maiorano et al. 2003;
Nysewander et al. 2003; Price \& Warren 2003; Rumyantsev,
Biryukov, \& Pozanenko 2003; Rumyantsev, Sergeeva \& Pozanenko
2003; Semkov 2003; von Braun, Garnavich \& Stanek 2003a, 2003b;
Zeh et al. 2003.}
\end{figure}


\begin{thebibliography}{99}
\bibitem{} Ando, M. et al. 2003a, GCN Circ. 1882
\bibitem{} Ando, M. et al. 2003b, GCN Circ. 1884
\bibitem{} Castor, J., McCray, R., \& Weaver, R. 1975, ApJ, 200, L107
\bibitem{} Chevalier, R. A., \& Imamura, J. N. 1983, ApJ, 270, 554
\bibitem{} Chevalier, R. A., \& Li, Z. Y. 1999, ApJ, 520, L29
\bibitem{} Chevalier, R. A., \& Li, Z. Y. 2000, ApJ, 536, 195
\bibitem{} Covino, S. et al. 2003, GCN Circ. 1909
\bibitem{} Dai, Z. G., \& Cheng, K. S. 2001, ApJ, 558, L109
\bibitem{} Dai, Z. G., \& Lu, T. 1998a, Phys. Rev. Lett., 81, 4301
\bibitem{} Dai, Z. G., \& Lu, T. 1998b, MNRAS, 298, 87
\bibitem{} Dai, Z. G., \& Lu, T. 2002, ApJ, 565, L87 (DL02)
\bibitem{} Fatkhullin, T. et al. 2003, GCN Circ. 1925
\bibitem{} Fox, D. W., Chen, H. W., \& Price, P. A. 2003, GCN Circ. 1879
\bibitem{} Frail, D. A. et al. 2001, ApJ, 562, L55
\bibitem{} Garcia-Segura, G. G., \& Mac Low, M. M. 1995, ApJ, 455, 145
\bibitem{} Garnavich, P., von Braun, K., \& Stanek, K. 2003, GCN Circ. 1885
\bibitem{} Greiner, J., Guenhter, E., Klose, S., \& Schwarz, R. 2003a, GCN Circ. 1886
\bibitem{} Greiner, J., Ries, C., Barwig, H., Fynbo, J., \& Klose, S. 2003b, GCN Circ. 1894
\bibitem{} Guarnieri, A. et al. 2003, GCN Circ. 1892
\bibitem{} Kobayashi, S., \& Zhang, B. 2003, ApJL, submitted (astro-ph/0304086)
\bibitem{} Kumar, P., \& Panaitescu, A. 2000, ApJ, 541, L51
\bibitem{} Li, Z. Y., \& Chevalier, R. A. 2003, ApJ, 589, L69
\bibitem{} Maiorano, E. et al. 2003, GCN Circ. 1933
\bibitem{} M\'esz\'aros, P., Rees, M. J., \& Wijers, R. A. M. J. 1998, ApJ, 499, 301
\bibitem{} Mirabal, N. et al. 2003, ApJ, submitted (astro-ph/0303616)
\bibitem{} Nakar, E., Piran, T., \& Granot, J. 2003, NewA, 8, 495
\bibitem{} Nysewander, M. C., Moran, J., Reichart, D., Henden, A., \& Schwartz, M. 2003,
                     GCN Circ. 1921
\bibitem{} Panaitescu, P., M\'esz\'aros, P., \& Rees, M. J. 1998, ApJ, 503, 314
\bibitem{} Pedersen, K., Fynbo, J., Hjorth, J., \& Watson, D. 2003, GCN Circ. 1924
\bibitem{} Price, P. A. et al. 2002, ApJ, 572, L51
\bibitem{} Price, P. A., Fox, D. W., \& Chen, H. W. 2003, GCN Circ. 1880
\bibitem{} Price, P. A., \& Warren, B. E. 2003, GCN Circ. 1890
\bibitem{} Ramirez-Ruiz, E., Dray, L. M., Madau, P., \& Tout, C. A. 2001, MNRAS, 327, 829
\bibitem{} Rhoads, J. 1999, ApJ, 525, 737
\bibitem{} Rumyantsev, V., Biryukov, V., \& Pozanenko, A. 2003, GCN Circ. 1908
\bibitem{} Rumyantsev, V., Sergeeva, L., \& Pozanenko, A. 2003, GCN Circ. 1929
\bibitem{} Sako, M., \& Fox, D. W. 2003, GCN Circ. 1928
\bibitem{} Sari, R., Piran, T., \& Halpern, J. P. 1999, ApJ, 519, L17
\bibitem{} Sari, R., Piran, T., \& Narayan, R. 1998, ApJ, 497, L17
\bibitem{} Semkov, E. 2003, GCN Circ. 1935
\bibitem{} Suzuki, M. et al. 2003, GCN Circ. 1888
\bibitem{} von Braun, K., Garnavich, P., \& Stanek, K. 2003a, GCN Circ. 1881
\bibitem{} von Braun, K., Garnavich, P., \& Stanek, K. 2003b, GCN Circ. 1902
\bibitem{} Wijers, R. A. M. J. 2001, in Gamma-Ray Bursts in the Afterglow Era,
           eds. E. Costa, F. Frontera \& J. Hjorth (Springer), 306
\bibitem{} Willis, A. J. 1991, in Wolf-Rayet Stars and Interrelations with Other Massive
           Stars in Galaxies, eds. K. A. van der Hulcht \& B. Hidayat
           (Dordrecht: Reidel), 265
\bibitem{} Wu, X. F., Dai, Z. G., Huang, Y. F., \& Lu, T. 2003, MNRAS, in press
           (astro-ph/0304110)
\bibitem{} Zeh, A., Klose, S., Greiner, J., Fynbo, J., \& Jakobsson, P. 2003,
           GCN Circ. 1898
\bibitem{} Zhang, B., \& M\'esz\'aros, P. 2002, ApJ, 566, 712
\end{thebibliography}
\end{document}